\documentclass[prb,aps,twocolumn,showpacs,superscriptaddress]{revtex4}
\usepackage{graphicx}
\begin{document}

\title{High magnetic field thermal-expansion and elastic properties of CeRhIn$_5$}

\author{V. F. Correa}
\affiliation{Los Alamos National Laboratory, Los Alamos, New Mexico 87545, USA}

\author{W. E. Okraku}
\affiliation{Occidental College, Los Angeles, California 90041, USA}

\author{J. B. Betts}
\affiliation{Los Alamos National Laboratory, Los Alamos, New Mexico 87545, USA}

\author{A. Migliori}
\affiliation{Los Alamos National Laboratory, Los Alamos, New Mexico 87545, USA}

\author{J. L. Sarrao}
\affiliation{Los Alamos National Laboratory, Los Alamos, New Mexico 87545, USA}

\author{A. H. Lacerda}
\affiliation{Los Alamos National Laboratory, Los Alamos, New Mexico 87545, USA}

\date{\today}


\begin{abstract}

We report high magnetic field thermal-expansion and magnetostriction results on 
CeRhIn$_5$ single crystals. 
Several transitions, both first and second order, are observed when the field is
applied perpendicular to the crystallographic c-axis.
The magnetic field dependence of the thermal-expansion coefficient above 15 K, 
where the magnetic correlations are negligible, can be explained supposing an 
almost pure $| \pm 5/2 \rangle$ ground state doublet, in apparent contradiction 
with neutron scattering experiments.
Although the spin-lattice interaction is relevant in this compound, the effect of 
the magnetic correlations on the elastic properties is relatively weak, as revealed 
by resonant ultrasound spectroscopy experiments.

\end{abstract}

\maketitle

\section{introduction}

The main characteristic of the heavy fermion (HF) systems is the strong 
interaction between the conduction electrons and the localized \textit{f}-electrons 
resulting in large electronic effective masses. This hybridization also gives rise 
to a non-magnetic Kondo singlet state or long-range magnetic order depending on 
the relative importance of the Kondo effect and the RKKY interaction.\cite{Hewson}

CeRhIn$_5$ is a member of the CeMIn$_5$ (M = Co, Rh, Ir) family.
\cite{Hegger,Petrovic1,Petrovic2} 
It crystallizes in the tetragonal HoCoGa$_5$ structure with alternating layers 
of magnetic CeIn$_3$ and non-magnetic MIn$_2$ along the c-axis.
CeRhIn$_5$ shows pressure induced superconductivity \cite{Hegger,Fisher} 
($T_c$ = 2.1 K at P = 16 kbar) and is the only member of the family that 
exhitis ambient pressure antiferromagnetism \cite{Hegger} ($T_N$ = 3.8 K).
The Ce magnetic moments, antiferromagnetically ordered, lie completely in the 
basal plane developing a helicoidal structure along the c-axis with a 
propagation vector \textbf{q} = (1/2,1/2,0.297) that is incommensurate with 
the atomic lattice.\cite{Curro,Bao1}
The absence of any magnetic order in CeIrIn$_5$ and CeCoIn$_5$ suggests that 
Rh plays a crucial role in establishing the magnetic structure. In fact, 
muon spin rotation experiments \cite{Schenck} suggest a small ordered magnetic 
moment at the Rh sites that might be responsible for the weak coupling of the 
Ce moments along the c-axis (via a superexchange mechanism) while the usual 
RKKY interaction accounts for the antiferromagnetic order within the basal 
plane.
The enhanced hybridization between the Ce-4\textit{f} electron and the conduction 
electrons would explain the heavy fermion superconducting state in CeIrIn$_5$ and 
CeCoIn$_5$. 

Another evidence of localized moment behavior is the crystal electric field 
(CEF) effect on the Ce-4\textit{f} electronic levels. Different experiments 
in CeRhIn$_5$ can be quite well understood in terms of CEF effects, although 
there are some differences in the proposed schemes: the main difference arising 
from the degree of admixture of the $| \pm 5/2 \rangle$ and $| \pm 3/2 \rangle$ 
levels in the ground state doublet. While thermal-expansion,\cite{Takeuchi} 
specific heat\cite{PagliusoB} and magnetic susceptibility\cite{Takeuchi,PagliusoB} 
data are explained supposing an almost pure $| \pm 5/2 \rangle$ ground state 
doublet, inelastic neutron scattering (INS) experiments\cite{Christianson1} 
clearly shows a peak at 23 meV that would be forbidden in such case.
Moreover, in a previous work\cite{Correa} in La doped Ce$_{0.6}$La$_{0.4}$RhIn$_5$ 
we showed that the magnetic field dependence of the thermal-expansion coefficient 
could only be explained by a pure $| \pm 5/2 \rangle$ ground state. In this case, 
the effect was atributed to changes in hybridization strength and chemical 
pressure due to the doping.  

On the other hand, an important Kondo compensation is also observed in 
CeRhIn$_5$ through neutron diffraction experiments\cite{Bao1} and corroborated 
by specific heat\cite{Hegger} and resistivity\cite{Christianson2} measurements, 
showing that there is a subtle competition bewteen the de-localized HF and the 
localized magnetic behaviors.

Despite the layered structure, the anisotropy displays a varying role. While 
resistivity\cite{Christianson2} and susceptibility\cite{Takeuchi} show little 
anisotropy, the magnetic and CEF contributions to the linear thermal-expansion
\cite{Takeuchi,Correa} present greater anisotropy. 
In the same way, the helicoidal magnetic structure suggests an important 
two-dimensional character. 
This is evident in specific heat\cite{Cornelius1} and magnetization\cite{Takeuchi,
Cornelius2} measurements where several field-induced transitions are observed 
only when the magnetic field is applied along the basal plane. 

Being that the spin-lattice coupling is a relevant issue in these Ce-based 
compounds,\cite{Takeuchi,Correa,Malinowski} in this work we investigate the 
magnetic effects on the volume and the elastic properties of CeRhIn$_5$ single 
crystals using high field linear thermal-expansion, magnetostriction and elastic 
constant measurements. 
Both first and second order transitions are observed in the magnetostriction and 
thermal-expansion coefficients giving rise to a field vs. temperature phase 
diagram similar to that obtained from specific heat experiments.\cite{Cornelius1}
However, the magnetic correlations seem to have a minor effect on the elastic 
constants.
The magnetic field dependence of the CEF contribution to the thermal-expansion 
cannot be explained by the levels scheme obtained from inelastic neutron 
scattering experiments.

\section{Results}

CeRhIn$_5$ single crystals were grown by the self flux technique.
The thermal-expansion experiments were performed using a capacitance dilatometer.
The elastic properties were studied using resonant ultrasound spectroscopy. 
\cite{Migliori} 

\begin{figure}[t]
\includegraphics[width=\columnwidth]{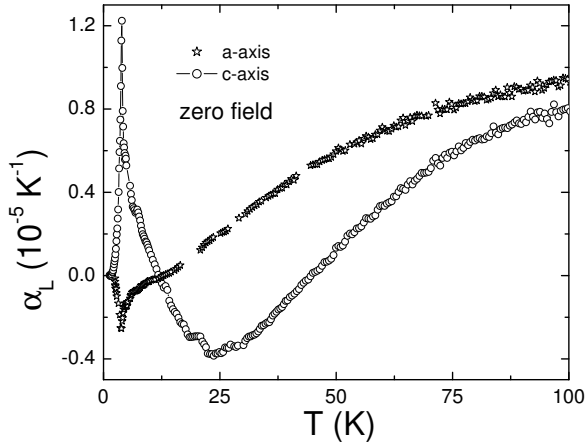}
\caption[]{Linear thermal-expansion vs temperature along the c-axis ($\alpha_c$) 
and perpendicular to it ($\alpha_{ab}$).}
\label{fig1}
\end{figure}

Figure \ref{fig1} shows the zero field linear thermal-expansion coefficient 
$\alpha = 1/L(dL/dT)$ along the c-axis of the crystal and perpendicular to it (i.e., 
along the ab-basal plane), $\alpha_c$ and $\alpha_{ab}$ respectively. 
As stated in previous works\cite{Takeuchi,Correa} two features can be distinguished: 
the antiferromagnetic transition is detected as a huge and anisotropic peak at low 
temperature, and the negative contribution to $\alpha_c$ around 25 K associated with 
crystal electric field (CEF) effects is observed.

\begin{figure}[t]
\includegraphics[width=\columnwidth]{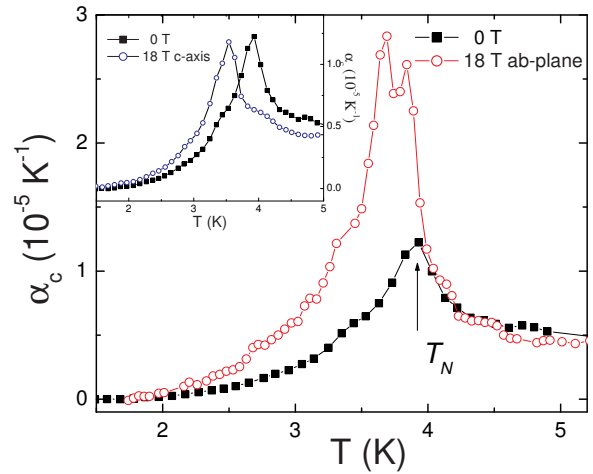}
\caption[]{Low temperature $\alpha_c$ showing the magnetic correlations associated 
peaks, for zero field and $B =$ 18 T applied along the ab-plane (main panel) and 
along the c-axis (inset).}
\label{fig2}
\end{figure}

The main panel of Fig. \ref{fig2} displays the low temperature $\alpha_c$ for zero 
field and $B =$ 18 T applied along the ab-plane. Clearly, the only peak associated 
with the N\"eel order at zero field splits into two peaks in the presence of a 
magnetic field. 
Also interestingly, the height of the peaks increases with field.
This must be compared with the observed behavior for fields applied along the c-axis,
shown in the inset of Fig. \ref{fig2}. Only one constant height peak at the 
magnetic transition is seen.

\begin{figure}[b]
\includegraphics[width=\columnwidth]{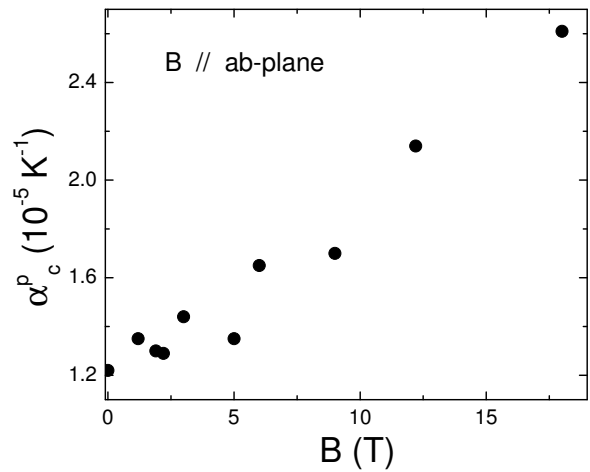}
\caption[]{Field dependence of $\alpha_c$ at $T_N$.}
\label{fig3}
\end{figure}

This is a consequence of the anisotropic nature of the magnetic structure. In the 
ordered phase, the magnetic moments lie in the basal plane. For low enough fields 
along the c-axis compared to the saturation field ($>$ 50 T, Ref. 
[\onlinecite{Takeuchi}]), the staggered magnetization is preserved with a small 
component of the magnetic moment in the direction of the field. So, the 
antiferromagnetic state is not destroyed and the magnetic field has a minor effect.  
However, the situation is different when the field is applied in the ab-plane.
In this case, a high enough field ($>$ 2 T, Ref. [\onlinecite{Takeuchi}]) breaks 
the antiferromagnetic symmetry fliping the moments in the direction of this external 
field and producing a rearrengement of the atomic orbitals which, in turn, gives rise 
to a larger volume change at the transition.
Naively, one could expect a height of the magnetic peak proportional to the field, as 
in fact is observed in Fig. \ref{fig3}.

\begin{figure}[t]
\includegraphics[width=\columnwidth]{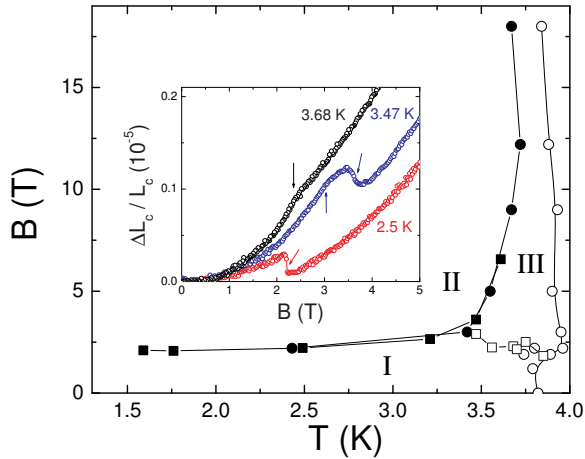}
\caption[]{Magnetic field vs. temperature phase diagram. Solid symbols are first 
order and open symbols are second order phase transitions. Squares points are from 
magnetostriction and circles are from thermal-expansion experiments. Inset: 
Linear magnetostricion showing first and second order transitions.}
\label{fig4}
\end{figure}

It is difficult to say something in CeRhIn$_5$ about the order of the phase transitions 
from thermal expansion measurements. A hint may be obtained from magnetostriction 
experiments. 
Typical results for fields in the ab-plane can be observed in the inset of Fig. 
\ref{fig4} where the field dependence of the c-axis is shown.
First and second order phase transitions can be clearly distingued in different 
temperature ranges.
Putting all this information together results in the magnetic field versus temperature 
phase diagram shown in the main panel of Fig. \ref{fig4}. There is a remarkable good 
agreement between thermal-expansion (open symbols) and magnetostriction (solid symbols) 
data. This phase diagram also has strong similarities with that obtained from 
specific heat experiments. \cite{Cornelius1}
According to Cornelius \textit{et al.},\cite{Cornelius1} phase I and II correspond to 
magnetic structures inconmensurate with the atomic lattice, while phase III corresponds 
to a conmensurate magnetic structure.  
Transitions between phases I and II, and bewteen phases II and III, are first order.

\begin{figure}[b]
\includegraphics[width=\columnwidth]{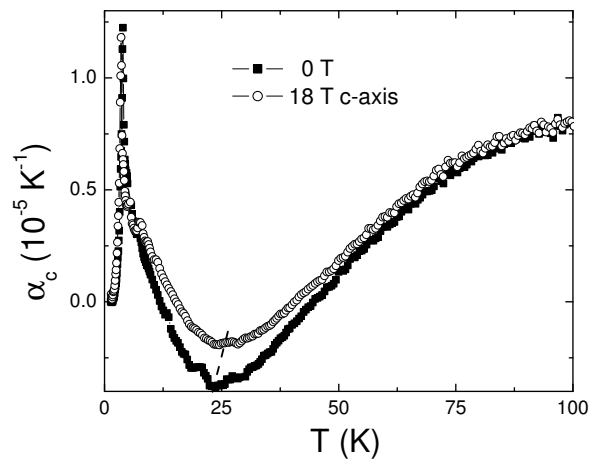}
\caption[]{Linear thermal-expansion $\alpha_c$ for zero field and $B =$ 18 T applied 
along the c-axis. The negative peak around 25 K, associated with CEF effects, moves 
towards higher T as a field is applied.}
\label{fig5}
\end{figure}

CEF effects are known to have relevant contributions to the specific heat, susceptibility 
and thermal-expansion of the 115 and the related 218 family.\cite{Takeuchi,PagliusoB,
Christianson1,Correa,Malinowski,Christianson3}
Figure \ref{fig5} shows $\alpha_c$ at zero field and 18 T along the c-axis in an 
extended temperature range. As previously observed\cite{Correa} in 
Ce$_{0.6}$La$_{0.4}$RhIn$_5$, the CEF-related negative peak around 25 K moves towards 
higher $T$ when the field is increased.
This is an anomalous behavior if we consider the CEF scheme obtained from inelastic 
neutron scattering measurements.\cite{Christianson1} According to these data, the 
CEF induced splitting of the Ce$^{3+}$ J = 5/2 multiplet results in a $\Gamma_7^{(2)} = 
 \sqrt{(1 - \eta^2)} \,\, | \pm 5/2 \rangle \, - \, \eta \,\, | \mp 3/2 \rangle$ ground 
state doublet with a mixing parameter $\eta =$ 0.6. 
However, as shown in Ref. [\onlinecite{Correa}], the field induced shift of the 
$\alpha_c$ minimum can only be explained supposing an almost pure $| \pm 5/2 \rangle$ 
ground state doublet ($\eta \sim$ 0). In that case, the result was attributed to 
chemical pressure and hybridization changes due to La doping.
But, what happens in the present case of non-doped CeRhIn$_5$? 
A peak in the INS spectra related to transitions from the ground state doublet to the 
second excited doublet (forbidden if $\eta =$ 0), is unambigously detected.
\cite{Christianson1}

The other contribution that we are perhaps neglecting is the influence of the Kondo 
interaction on the CEF effects and vice versa.
The Kondo scattering is reduced by the CEF splitting giving rise to an 
effective Kondo temperature $T_K^{eff}$ for the ground doublet state that can be much 
lower than the Kondo temperature $T_K$ for the whole $J =$ 5/2 multiplet.\cite{PagliusoB,
Suzuki}
On the other hand, as a result of the hybridization between the conduction and the 
\textit{f}-electrons, the CEF levels become dispersive bands whose width is $T_K^{eff}$   
(Ref. [\onlinecite{Christianson1}]). 

According to the CEF scheme from INS results, a magnetic field splits the three doublets 
and mixes together the ground state and the first excited doublet. 
For $B =$ 18 T, the separation between the two lowest singlet states is $\Delta_1^S$ (18 T) 
$\sim$ 21 K. Hence, for the whole field range we can access (up to 18 T),
$\Delta_1^S (B) < T_K^{eff} \sim$ 25 K (Ref. [\onlinecite{Christianson1}]), and both 
states overlap.  
Moreover, the expected value of the magnetic moment 
$g \mu_B \langle J_z \rangle$ along $B$ (= 0.92 $\mu_B$ for zero field) for both  
the ground and first excited singlet states are amazingly different: 1.52 and 0.02 
$\mu_B$, respectively. 
As the Kondo effect is an interaction between the free electron and the localized 
magnetic moments, different hybridization can be expected for different magnitudes 
of the magnetic moment. 

In fact, a good fit to the specific heat data needs not only a CEF contribution but also a 
Kondo contribution, too.\cite{PagliusoB,Christianson1} So, it is reasonable to 
think that the same would be necessary for the thermal-expansion data.
In this way, it must be stressed that fits to $\alpha$ disregarding the Kondo effect,
\cite{Takeuchi} lead to similar results as ours: an almost pure $| \pm 5/2 \rangle$ 
ground state doublet.

\begin{figure}[t]
\includegraphics[width=\columnwidth]{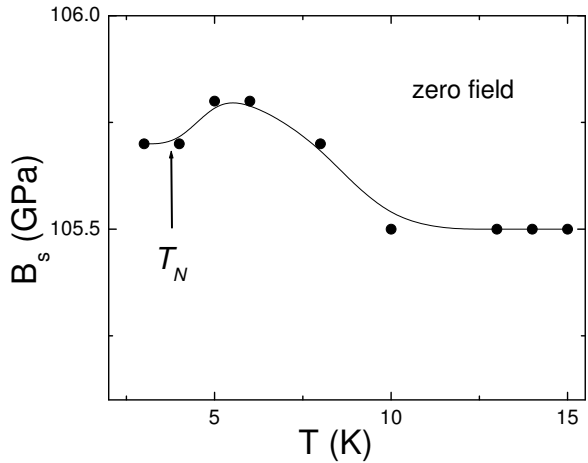}
\caption[]{Temperature dependence of the adiabatic bulk modulus.}
\label{fig6}
\end{figure}

Finally, we study the effect of the magnetic correlations on the elastic properties.
Figure \ref{fig6} shows the temperature dependence of the adiabatic bulk modulus  
$B_S$ at zero field. Only a very subtle effect can be observed across $T_N$. 
The same results are obtained for any elastic constant across any phase boundary 
displayed in Fig. \ref{fig4}. However, a small increase in 
$B_S$ is observed below 10 K, that is roughly the temperature up to which the 
antiferromagnetic fluctuations are detected.\cite{Bao3D} 
It leads us to the conclusion that although the magnetic correlations have an 
important effect on the lattice dimensions (a relevant spin-lattice coupling), 
as shown by thermal-expansion and magnetostriction measurements, they have a 
less important effect on the harmonic part of the interatomic potential energy, 
as evidenced by the elastic constants.

\section{Conclusions}

Thermal-expansion, magnetostriction and elastic constant measurements were 
performed in CeRhIn$_5$ single crystals.
A rich magnetic field vs. temperature phase diagram is observed when the field 
is applied along the ab-plane, in good agreement with specific heat experiments.
\cite{Cornelius1}
Both first and second order transtions can be clearly distinguished.
The magnetic field dependence of the CEF contribution to the thermal-expansion 
cannot be explained by the levels scheme obtained from inelastic neutron 
scattering experiments, suggesting that perhaps the effect of the Kondo 
interaction on the thermal-expansion should be considered.
The magnetic correlations show no major effect on the elastic properties, the 
most relevant effect being a small increase in the adiabatic bulk modulus 
below 10 K, presumedly associated with the antiferromagnetic fluctuations.

\section{Acknowledgments}

We gratefully acknowledge J. M. Lawrence for helpful discussions. Work at the NHMFL 
was performed under the auspices of the National Science Foundation, the State of 
Florida and the US Department of Energy.

\end{document}